\documentclass[11pt,twoside]{article}
\usepackage{amsmath}	
\usepackage{asp2010}
\usepackage{natbib}

\def\vec#1{\ensuremath{\mathbf{#1}}}

\def\alf{Alfv\'en}

\resetcounters

\begin{document}

\title{Parallel-cascade-based mechanisms for heating solar coronal loops: test against observations}
\author{Bo Li$^1$, Haixia Xie$^1$, Xing Li$^2$, and Li-Dong Xia$^1$}

\affil{$^1$Shandong Provincial Key Laboratory of Optical Astronomy \& Solar-Terrestrial Environment,
   School of Space Science and Physics,
   Shandong University at Weihai, Weihai 264209, China}
\affil{$^2$Institute of Mathematics \& Physics, Aberystwyth University, Aberystwyth SY23 3BZ, UK}

\begin{abstract}
The heating of solar coronal loops is at the center of the problem of coronal heating.
Given that the origin of the fast solar wind has been tracked down to atmospheric layers with transition region
    or even chromospheric temperatures, it is worthy attempting to address whether the mechanisms proposed to provide
    the basal heating of the solar wind apply to coronal loops as well.
We extend the loop studies based on a classical parallel-cascade scenario originally proposed in the solar wind context
    by considering the effects of loop expansion, and perform a parametric study to directly
    contrast the computed loop densities and electron temperatures with those measured by
    TRACE and YOHKOH/SXT.
This comparison yields that with the wave amplitudes observationally constrained by SUMER measurements,
    while the computed loops may account for a significant fraction of SXT loops,
    they seem too hot when compared with TRACE loops.
Lowering the wave amplitudes does not solve this discrepancy, introducing magnetic twist will make the
    comparison even less desirable.
We conclude that the nanoflare heating scenario better explains ultraviolet loops,
    while turbulence-based steady heating mechanisms may be at work in heating a fraction of 
    soft X-ray loops.
\end{abstract}

\section{Modeling solar coronal loops}

How the solar corona is heated to multi-million degrees of Kelvin remains a topic of
    intensive study~\citep{2006SoPh..234...41K,2012RSPTA.370.3217P}.
Due to their higher demand of energy flux consumption~\citep[e.g.,][]{1977ARA&A..15..363W}, 
    loop structures -- the magnetically closed part of the corona --
    receive more attention than coronal holes -- their magnetically open counterpart.
Conventionally loop heating mechanisms are grouped into two categories: DC ones that involve the dissipation of the energy
    of the magnetic field stressed by supergranular motions most likely via magnetic reconnections
    at small scale current sheets, and AC ones that involve the deposition of energy that ultimately derives
    also from supergranular motions but is transported as waves.

Actually the fast solar wind that emanates from coronal holes also requires a basal heating.
Their origin, originally attributed to the vaguely defined ``coronal base'' where the temperature
    has reached a million degree, has been observationally tracked down to
    the atmospheric layers above chromospheric network~\citep{1999Sci...283..810H,2005Sci...308..519T}.
Not only supplying the required mass, the chromospheric activities may also provide the required energy for heating and transporting
    the materials from the upper chromosphere to the corona~\citep{2011ApJ...727....7M}.
Stimulated by these measurements, modern fluid models of the solar wind tend to place the inflow boundary at
    the transition region or the upper chromosphere or even at photospheric levels~\citep[e.g.,][]{2012SSRv..172..145C}.
To provide the needed heating for the nascent fast solar wind, modern models tend to use 
    either observationally based empirical heating functions, or the heating rates due to the
    dissipation of various waves via, say, turbulent means.
    
It seems natural but in fact rather rare to see coronal loop models heated by mechanisms originally devised for
    heating nascent solar winds.
The available ones are mainly based on the resonant interactions between protons
    and ion-cyclotron waves, which were designed in the solar wind context to naturally account for 
    the temperature measurements above coronal holes, 
    especially the inferred significant ion temperature anisotropy~\citep[e.g.,][]{2002JGRA..107.1147H}.
The needed ion-cyclotron waves may be generated either by a turbulent parallel cascade from low-frequency \alf\ waves
    emitted by the Sun~\citep{2003ApJ...598L.125L, 2005A&A...435.1159O}, or directly by small-scale magnetic reconnection events at
    chromospheric network~\citep{2008ApJ...676.1346B}.
While by construction the waves heat protons only, electrons may readily receive part of the heating via frequent 
    collisions with protons given the high loop densities.
These ion-cyclotron resonance based mechanisms were shown to be able to produce a million-degree loop with realistic
    densities.
A salient feature of these models is that, when only unidirectional waves are introduced, the heating is generally not symmetric
    with respect to the looptop, resulting in substantial loop flows.
These flows are essential in enhancing the loop densities relative to hydrostatic expectations.
In parallel-cascade based models, it was also shown that the ponderomotive force density associated with the \alf\ waves
    plays an important role in the loop dynamics, especially close to the loop ends~\citep{2003ApJ...598L.125L}.
When magnetic twist is introduced, the electron temperature may be significantly enhanced due to the projection
    effect~\citep{2006RSPTA.364..533L}.

In contrast to the extensive attempts in the loop community to directly contrast model computations with 
    observations~\citep[e.g.,][]{2003ApJ...587..439W},
    the loop models using solar wind heating mechanisms have not been tested against observations.
Of particular interest would be the loop density and temperature, which are the most frequently measured parameters.
In this presentation we will present a preliminary study along this line of thinking.
Specifically, the data that will be compared with are obtained by the ultraviolet instruments onboard TRACE
    and the X-ray instrument SXT onboard YOHKOH as compiled in~\citet{2003ApJ...587..439W}.
We note that the filter ratio technique in deducing the temperatures may be subject to considerable uncertainty,
    however, let us only mention the limitations of the loop models here.
The models are based on the parallel-cascade scenario where the \alf\ waves are injected
    at one loop end, and via a parallel cascade the wave energy is transferred to the ion-cyclotron range
    and therefore readily picked up by protons via proton cyclotron resonance~\citep{2002JGRA..107.1147H,2003ApJ...598L.125L}.
By using unidirectional waves described by a WKB-like equation supplemented with dissipation, 
    we assume that the backward propagating waves, which are essential in generating
    any MHD cascade, do not contribute significantly to the energy flux density.
In this sense the wave frequencies are higher than the \alf\ speed divided by its characteristic spatial scale.
For the computed values it was found that this frequency would be of the order of one hundred Hertz, which seems high
    but consistent with the estimated frequencies of the \alf\ waves launched
    by chromospheric magnetic reconnections~\citep{1999ApJ...521..451S}.
In future a more self-consistent treatment of bi-directional waves and their dissipation due to mutual coupling
    should be pursued, say, in the manner proposed by~\citet{2013ApJ...764...23S}.

\section{Model description}
\label{sec_model}
We approximate coronal loops as a semi-circular torus with length $L$ and cross-sectional area $a$.
The loop magnetic field $B$ as a function of arclength $l$, measured from one loop footpoint along the axis,
    is related to $a$ via $B\propto 1/a$.
The loop material consists of electrons ($e$) and protons ($p$), and each species $s$ ($s = e, p$) is
    characterized by its number density $n_s$, mass density $\rho_s = n_s m_s$, temperature $T_s$, velocity $\vec{v}_s$,
    and partial pressure $p_s = n_s k_B T_s$ with $k_B$ being the Boltzmann constant.
Quasi-neutrality ($n_e = n_p = n$) and quasi-zero-current ($\vec{v}_e = \vec{v}_p = \vec{v}$) are assumed.
Only monolithic loops in steady state are considered, i.e., $\partial/\partial t=0$, 
    and the variation in the direction perpendicular to the loop axis is neglected.
With electron inertia further neglected, the standard two-fluid MHD equations are then projected along the loop axis, rendering $l$
    the only independent variable.
The governing equations read ({for more details, please see~\citeauthor{2003ApJ...598L.125L}~\citeyear{2003ApJ...598L.125L}})    
\begin{align}
& (n v a)' =0, 	\label{eq_density} \\
& vv'       =-\frac{(p_e+p_p)'}{\rho}
            - g_\parallel +\frac{F}{\rho} ,
	\label{eq_momen} \\
& v \left(T_e\right)'
         + \frac{(\gamma-1)T_e \left(a v\right)' }{a}
         =\frac{\gamma-1}{n k_B a}\left(a \kappa_{e0} T_e^{5/2} T_e'\right)'
&         -2\nu_{pe} (T_e-T_p) - \frac{\gamma-1}{n k_B} L_{\mathrm{rad}} , \label{eq_Te} \\
& v \left(T_p\right)'
         + \frac{(\gamma-1)T_p \left(a v\right)' }{a}
         =\frac{\gamma-1}{n k_B a}\left(a \kappa_{p0} T_p^{5/2} T_p'\right)' 
&       + 2\nu_{pe} (T_e-T_p) + \frac{\gamma-1}{n k_B} Q_{\mathrm{wav}} , \label{eq_Tp} 
\end{align}
in which the prime $'$ denotes the differentiation with respect to $l$, and $\gamma=5/3$ is the adiabatic index.
Furthermore, $\rho=\rho_p$ is the total mass density, and 
    $g_\parallel$ denotes the gravitational acceleration corrected for loop curvature.
The Coulomb collision rate $\nu_{pe}$ 
    is evaluated by using a Coulomb logarithm of $23$.
The electron energy loss is denoted by $\cal{L}_{\rm rad}$,
    and we adopt the standard parametrization by~\citet{1978ApJ...220..643R} for an optically thin medium.
Besides, $\kappa_{e0} = 7.8\times 10^{-7}$ and $\kappa_{p0} = 3.2\times 10^{-8}$ represent the Spitzer values for the species 
    thermal conductivities (cgs units will be used throughout).
By construction the energy deposition $Q_{\mathrm{wav}}$ due to waves goes entirely to
    heating protons, and is related to the wave evolution via
\begin{align}
 \frac{\left(a F_w\right)'}{a} + v F = -Q_{\mathrm{wav}} , \label{eq_wave}
\end{align}
    where $F=-p_w'$ and $F_w$ are the wave force and energy flux densities, respectively.
Consistent with previous solar wind models, here $Q_{\mathrm{wav}}$ is assumed to follow a Kolmogorov rate, 
    $Q_{\mathrm{wav}} = \rho \xi^3/L_{\mathrm{corr}}$, where $\xi$ denotes the wave amplitude, 
    and $L_{\mathrm{corr}}$ denotes the correlation length associated with turbulent heating.
As conventionally assumed, $L_{\mathrm{corr}}$ is proportional to $1/\sqrt{B}$~\citep{2002JGRA..107.1147H}.

Now we need to specify the axial distribution of the loop magnetic field strength $B(l)$, which is assumed to be symmetric
    about the looptop ($l=L/2$).
We distinguish between two profiles, in one of which $B\equiv 60$~G is uniform  and in the other it decreases from 240~G at
    loop ends to 60~G at looptop with the specific profile parametrized following the measurements
    of the loop cross-sectional area, deduced from the width of supergranular network at a range of Ultraviolet lines formed
    at different temperatures~\citep{2006ApJ...647L.183A}.

The following boundary conditions are used.
At both ends ($l=0$ and $L$), the number density $n$ and speed $v$ are allowed to change freely, mimicking
     the filling and draining of loop materials due to coupling with the underlying denser layer.
However, both electron and proton temperatures are fixed at $2\times 10^4$~K, corresponding to 
     the top of the chromosphere.
The wave amplitude $\xi_0$ at the driving end ($l=0$) where the waves enter the loop
     is $10$~km/s, in line with the SUMER measurements with linewidths~\citep{1998ApJ...505..957C},
     but is allowed to vary freely at the outflowing end ($l=L$).
As such, a solution is uniquely determined once one specifies the loop length $L$ and the correlation
     length  $l_0$ at the driving end, enabling us to perform a parametric study on the range the loops parameters may span
     at a range of loop lengths.
     
A description of the axial profiles of loop parameters is
     necessary~\citep[for details, see][]{2005A&A...435.1159O}.
At a given $L$, for all the chosen $l_0$ the electron density $n$ decreases from some chromospheric value at 
      the driving end, attains a minimum, and then increases
      again towards a chromospheric value at the other end.
The associated proton speed $v$ and electron temperature $T_e$ exhibit an opposite fashion.
Their specific profiles critically depend on the choice of $l_0$, 
     with the tendency being that
     when $l_0$ increases, the wave heating becomes more uniform and the loop becomes less dynamic, i.e.,
     the maximum speed decreases.
If $l_0$ is larger than some critical value, the loop becomes static and there is practically no flow at all.
If on the contrary $l_0$ is smaller than a critical value, the loop is so dynamic that a slow shock develops at one end.
The shocked solutions may be important on theoretical grounds, but their observational detection in coronal loops
     has not been reported.
We therefore are left with a range of $l_0$, only in which the solutions are observationally accessible.
Varying $l_0$ in this range, we find for any given $L$ the ranges for the minimal electron density $n_{\mathrm{Min}}$
     and maximum electron temperature $T_{\mathrm{Max}}$, which are then compared with observations.
     
\section{Comparison of model results with SXT and TRACE observations}
\label{sec_model_res}

\begin{figure}
\centering
\includegraphics[width=0.8\textwidth]{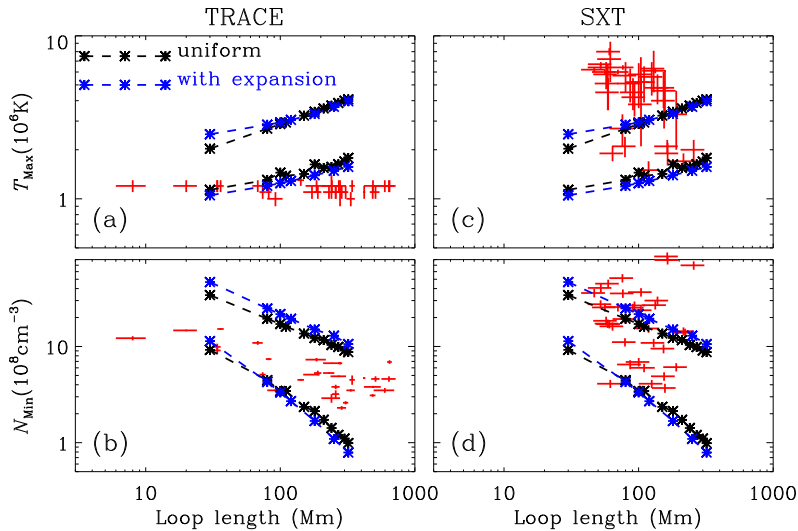}
\caption{
Comparison with measurements of loop parameters from models based on parallel cascade of \alf\ waves.
Panels (a) and (b) display the computed range of the electron temperature maximum $T_{\mathrm{Max}}$
    as a function of looplength.
Likewise, panels (c) and (d) give the corresponding distribution of the ranges of the minimum electron density 
    $N_{\mathrm{Min}}$.
The black dashed lines represent model computations where the loop cross-sectional area does not vary with distance,
    while the blue ones are for models where the loop experience some lateral expansion.
Besides, the red crosses in the left (right) column display the parameters of the loops measured with TRACE (YOHKOH/SXT).
}
\label{fig_loop}
\end{figure}

Figure~\ref{fig_loop} presents the results from this parametric survey, which displays the computed ranges for $n_{\mathrm{Min}}$
     and $T_{\mathrm{Max}}$ as a function of looplength $L$.
The red crosses are the measured values for the TRACE (left column)
     and YOHKOH/SXT (right) loops, read from Tables 1 and 2 compiled by~\citet{2003ApJ...587..439W}. 
The black dashed curves are for the computations where the loop magnetic field is uniform,
     whereas the blue curves are for the case where loops experience some expansion.
Specific computations are represented by the asterisks.
It is clear from the figure that while varying the loop cross-section may drastically change the axial profiles
     of the loop parameters (not shown), the ranges the electron densities and temperatures may span 
     are not substantially different: the ranges are slightly broader in the expanding case.
From the left column, it is clear that while the electron densities measured by TRACE are reproduced remarkably well,
     the computed loops are too hot compared with measurements.
In fact, among the 22 loops with lengths ranging from 30 to 300~Mm, literally all the measured values lie in
     the computed $n_{\mathrm{Min}}$ ranges, whereas only 3 lie in the computed $T_{\mathrm{Max}}$ ranges. 
An intuitive idea would that if we decrease the wave amplitude, this comparison would be more desirable, but this turns out not
     to be the case: lowering $\xi_0$ to 7~km/s, we found the computed loop temperatures are still too high.
If introducing magnetic twist, which is often observed to be present in coronal loops, we would find that the loops are even
     hotter~\citep{2006RSPTA.364..533L}.     
So we conclude here that at this level of sophistication, the parallel-cascade based mechanisms cannot explain the EUV loops, whose
     flat distribution of temperatures just above 1~MK may be better explained by the impulsive heating scenarios,
     e.g., the nanoflare approach.
     
The computed loop parameters compare more favorably with SXT measurements.
While not perfectly reproduced, among the 47 loops measured, 10 loops lie in the computed  $T_{\mathrm{Max}}$ ranges,
     with an additional 3 being possible when the measurement uncertainty is considered.
As for the loop densities, 20 of the measured values lie in the computed ranges.
From this we conclude that the steady heating model based on this parallel cascade scenario may account for a substantial fraction
     of the soft X-ray loops.
Interestingly, there is observational evidence that the X-ray emitting Active Region cores may last 
     hours, thereby partly lending support to some steady heating. 

\section{Summary}
\label{sec_disc}
The problem of coronal heating largely concerns the question of how to heat the magnetically closed part of
    the Sun -- coronal loops -- to multi-million degrees of Kelvin.
However, there is ample evidence that the solar wind, at least the fast streams, originates from the atmospheric layers
    as low as the top of the chromosphere, and therefore has to undergo
    some basal heating to bring their temperature to a million degree as well.
In this sense it is worth examining whether the mechanisms designed for heating the nascent fast solar wind
    can be also applied to coronal loop heating.
This was undertaken by~\citet{2003ApJ...598L.125L, 2005A&A...435.1159O, 2006RSPTA.364..533L, 2008ApJ...676.1346B}.
Somehow these attempts still lack a rigorous observational test: there is neither an attempt to reproduce 
    a particular observed loop, nor a study to examine whether the proposed mechanisms can reproduce the observed loop ensembles
    with different instruments.
We present a preliminary attempt that falls in the second category, examining the applicability of parallel-cascade based mechanisms where
    ion-cyclotron resonance plays the central role.
However, we found that with the observationally constrained wave amplitudes, this mechanism cannot reproduce the TRACE loops, for the computed 
    loop temperatures are always higher than observed.
Nonetheless, the computed loop densities and temperatures can reproduce a substantial fraction of the SXT loops.
Given that the solar wind studies have accumulated a considerable set of mechanisms, a serious need exists to test 
    their applicability to loop heating in 
    a systematic manner against observations, such as was conducted in the present study.

{Before closing, we note that the conclusions drawn here apply only to the parallel-cascade scenario.
It remains to be seen whether perpendicular-cascade-based mechanisms, now intensively pursued in the solar
    wind community\citep[e.g.,][]{2011ApJ...743..197C,2013ASPC..474..153L},
    can reproduce the ultraviolet observations of coronal loops.
Such a study, however, is left for a future publication.}

\acknowledgements
This research is supported by the 973 program 2012CB825601, the National Natural Science Foundation of China (40904047,
  41174154, 41274176, and 41274178), and by the Ministry of Education of China (20110131110058 and NCET-11-0305).

\bibliographystyle{asp2010}
\bibliography{Li_etal}

\end{document}